\documentstyle[preprint,aps,epsfig]{revtex}
\tighten
\begin{document}
\draft
\preprint{Preprint Numbers: \parbox[t]{45mm}{ANL-PHY-8432-TH-96\\ 
                                MPG-VT-UR 83/96\\
                        nucl-th/9606006}}
\title{Continuum study of deconfinement at finite temperature}
\author{Axel Bender\footnotemark[1], David Blaschke\footnotemark[2], Yuri
Kalinovsky\footnotemark[2]\footnotemark[3] and Craig
D. Roberts\footnotemark[1]
\vspace*{0.3\baselineskip}} 
\address{\footnotemark[1]
Physics Division, Bldg. 203, Argonne National Laboratory,
Argonne, Illinois 60439-4843\vspace*{0.3\baselineskip}}
\address{\footnotemark[2]
MPG Arbeitsgruppe ``Theoretische
	 Vielteilchenphysik''\\
	 Universit\"at Rostock, D-18051 Rostock, Germany
        \vspace*{0.3\baselineskip}}
\address{\footnotemark[3]
Bogoliubov Laboratory of Theoretical Physics,\\
	   Joint Institute for Nuclear Research,
	   141980 Dubna, Russia\vspace*{0.3\baselineskip}}
\maketitle
\begin{abstract}
Deconfinement and chiral symmetry restoration are explored in a confining,
renormalisable, Dyson-Schwinger equation model of two-flavour QCD.  An order
parameter for deconfinement is introduced and used to establish that, in the
chiral limit, deconfinement and chiral symmetry restoration are coincident at
$T_c\approx 150\,$MeV.  The transitions are second order and each has the
same critical exponent: $\beta\approx 0.3$.  The deconfinement transition is
found to exhibit sensitivity to the current-quark mass.  $f_\pi$ and $m_\pi$
change by no more than 10\% for $T<0.7\,T_c$, however, as $T\to T_c$, thermal
fluctuations cause the pion bound state contribution to the four-point
quark-antiquark correlation function to disappear.
\end{abstract}
\pacs{Pacs Numbers: 11.10.Wx, 12.38.Aw, 11.30.Rd, 24.85.+p, 11.15.Tk}
%
%
The Dyson-Schwinger equations [DSEs] provide a nonperturbative,
renormalisable, continuum framework for analysing quantum field theories.  An
important example is the fermion DSE, which has proven useful in the study of
confinement and dynamical chiral symmetry breaking [DCSB]~\cite{M95}.  The
DSEs form a tower of coupled equations, which must be truncated to arrive at
a tractable problem.  Truncations that preserve the global symmetries of a
field theory are easy to implement.  Preservation of a gauge symmetry is more
difficult but there is progress in that direction~\cite{BP94}.  The
elementary quantities are the Schwinger functions whose analytic properties
provide information about confinement and DCSB; e.g., the absence of a
Lehmann representation for the two-point dressed-gluon Schwinger function
(gluon propagator) entails the absence of asymptotic gluon states; i.e.,
gluon confinement.  The approach is reviewed in Ref.~\cite{DSErev}.

The finite temperature properties of QCD are important in astrophysics and
cosmology, and may be explored in a future-generation heavy-ion accelerator
programme.  Theoretical tools that can be employed reliably in the
nonperturbative study of deconfinement and chiral symmetry restoration are
therefore valuable.  In Ref.~\cite{FR96} a one-parameter model dressed-gluon
propagator was proposed, which provided a good description of $\pi$ and
$\rho$-meson observables.  The single parameter in this model is a mass
scale, $m_t$, that marks the point where the nonperturbative, infrared
enhancement found in gluon DSE studies~\cite{IRsing} becomes dominant.  The
model gluon propagator has no Lehmann representation.  The calculated quark
propagator also has no Lehmann representation; therefore both the gluon and
quark are confined.  With $m_t$ fixed at $T=0$, one has a renormalisable
DSE-model of QCD, which manifests both confinement and DCSB, whose finite
temperature behaviour may provide insight into the finite temperature
properties of QCD.

The DSE for the renormalised dressed-quark propagator at finite temperature
involves a sum over Matsubara frequencies.  In nonperturbative DSE studies,
where the analytic structure of the dressed-quark propagator is calculated
rather than assumed, it is necessary to perform this sum numerically because
the usual analytic methods of evaluating it rely upon the quark propagator
having a Lehmann representation, which may not be the case for a confined
quark.  The absence of a Lehmann representation also complicates, if not
precludes, a real-time formulation of the finite temperature theory.

Herein we employ a Euclidean space formulation with
$\{\gamma_\mu,\gamma_\nu\}= 2\delta_{\mu\nu}$,
$\gamma_\mu=\gamma_\mu^\dagger$, $\beta\equiv \gamma_4$, $\gamma\cdot p
\equiv \sum_{i=1}^3\,\gamma_i\, p_i$, $\omega_n \equiv (2 n + 1)\pi T$ and
$\Omega_n \equiv 2 n \pi T$, where $T$ is the temperature .  The DSE for the
renormalised quark propagator, $S \equiv [- i\gamma\cdot p \sigma_A(p,\omega_n)
- i\beta\,\omega_n\sigma_C(p,\omega_n) + \sigma_B(p,\omega_n)]$, is
\begin{eqnarray}
\nonumber
\lefteqn{S^{-1}(p,\omega_n) = }\\
&&Z_2^A \,i\gamma\cdot p + Z_2 \, (i\beta\,\omega_n + m_{\rm bm})\,
        + \Sigma^\prime(p,\omega_n),
\label{qDSE}
\end{eqnarray}
$m_{\rm bm}$ is the bare mass and the regularised self energy is
\begin{eqnarray}
\nonumber
\lefteqn{\Sigma^\prime(p,\omega_n) =}\\
&& i\gamma\cdot p\,\Sigma_A^\prime(p,\omega_n)
+ i\beta\,\omega_n\,\Sigma_C^\prime(p,\omega_n) + 
\Sigma_B^\prime(p,\omega_n)\; ,
\end{eqnarray}
with
\begin{eqnarray}
\nonumber
\lefteqn{\Sigma_{\cal F}^\prime(p,\omega_n) =\int_{m,q}^{\bar\Lambda}\,
\case{4}{3}\,g^2\,D_{\mu\nu}(p-q,\omega_n-\omega_m)}\\
&&
\times \case{1}{4}{\rm tr}\left[{\cal P}_{\cal F}
        \gamma_\mu S(q,\omega_m)\Gamma_\nu(q,\omega_m;p,\omega_n)\right]\,
\label{regself}
\end{eqnarray}
where ${\cal F}=A,B,C$; ${\cal P}_A\equiv -(Z_1^A/p^2)i\gamma\cdot p$, ${\cal
P}_B\equiv Z_1 $, ${\cal P}_C\equiv -(Z_1/\omega_n)i\beta$, and
$\int_{m,q}^{\bar\Lambda}\equiv\, T
\,\sum_{m=-\infty}^\infty\,\int^{\bar\Lambda}\frac{d^3q}{(2\pi)^3}$.  In
Eq.~(\ref{regself}), $\Gamma_\nu(q,\omega_m;p,\omega_n)$ is the renormalised
dressed quark-gluon vertex and $ D_{\mu\nu}(p,\Omega_n)$ is the renormalised
dressed-gluon propagator.  One can also write
\begin{equation}
S^{-1}(p,\omega_n) \equiv i\gamma\cdot p \,A(p,\omega_n) 
+ i\beta\,\omega_n \,C(p,\omega_n)
        + B(p,\omega_n).
\end{equation}

In renormalising we require that
\begin{equation}
\label{subren}
\left.S^{-1}(p,\omega_0)\right|_{p^2+\omega_0^2=\mu^2} = 
        i\gamma\cdot p + i\beta\,\omega_0 + m_R\;,
\end{equation}
which entails that the renormalisation constants are given by:
$Z_2^A(\mu,\bar\Lambda) = 1- \Sigma_A^\prime(\mu,\omega_0;{\bar\Lambda})$,
$Z_2(\mu,\bar\Lambda) = 1- \Sigma_C^\prime(\mu,\omega_0;{\bar\Lambda})$,
$m_R(\mu) = Z_2 m_{\rm bm}({\bar\Lambda}^2) +
\Sigma_B^\prime(\mu,\omega_0;{\bar\Lambda})$, and the renormalised self
energies are
\begin{equation}
\begin{array}{rcl}
{\cal F}(p,\omega_n;\mu) & = & 
\kappa_{\cal F} + \Sigma_{\cal F}^\prime(p,\omega_n;{\bar\Lambda})
    - \Sigma_{\cal F}^\prime(p,\omega_0;{\bar\Lambda})\,,
\end{array}
\end{equation}
${\cal F}=A,B,C$, $\kappa_A = 1 = \kappa_C$ and $\kappa_B=m_R(\mu)$.

It is invalid to neglect $A(p,\omega_n;\mu)$ and $C(p,\omega_n;\mu)$ and
their dependence on their arguments.  In studying confinement the
$(p,\omega_n)$-dependence of $A$ and $C$ is qualitatively important since it
can conspire with that of $B$ to eliminate free-particle poles in the quark
propagator~\cite{BRW92}.  Further, all calculated observables are
quantitatively sensitive to $A$ and $C$; e.g., in the present study we find
$A(0,\omega_0) \sim 1.5$ and $C(0,\omega_0) \sim 1.5-1.8$ (increasing with
$T$) and they return to $1$ at the renormalisation point - neglecting this
variation entails $f_\pi \to 2 f_\pi$, which one may not be able to
compensate by readjusting a model's parameters.

To study the chiral limit one must solve the DSE with $m_R(\mu)=0$, {\it not}
$m_{\rm bm}=0$.  We denote solution functions obtained in the chiral limit by a
subscript ``$0$''; e.g., $B_0(p,\omega_n;\mu)$ and
$\sigma_{B_0}(p,\omega_n;\mu)$.

The finite-$T$ gluon propagator is
\begin{equation}
g^2 D_{\mu\nu}(p,\Omega) = 
P_{\mu\nu}^L(p,\Omega) \Delta_F(p,\Omega) + 
P_{\mu\nu}^T(p) \Delta_G(p,\Omega) 
\end{equation}
in Landau gauge, where
\begin{eqnarray}
P_{\mu\nu}^T(p) & \equiv &\left\{
\begin{array}{c}
0; \; \mu\;{\rm and/or} \;\nu = 4,\\
\displaystyle
\delta_{ij} - \frac{p_i p_j}{p^2}; \; \mu,\nu=1,2,3
\end{array}\right.
\end{eqnarray}
with $P_{\mu\nu}^T(p) + P_{\mu\nu}^L(p,p_4) = \delta_{\mu\nu}- p_\mu
p_\nu/{\sum_{\alpha=1}^4 \,p_\alpha p_\alpha}$; $\mu,\nu= 1,\ldots, 4$.  A
``Debye-mass'' for the gluon appears as a $T$-dependent contribution to
$\Delta_F$.

We employ a parameter-free finite-$T$ extension of the model dressed-gluon
propagator of Ref.~\cite{FR96}: $\Delta_F(p,\Omega) \equiv {\cal
D}(p,\Omega;m_D)$ and $\Delta_G(p,\Omega) \equiv {\cal D}(p,\Omega;0)$;
\begin{eqnarray}
\nonumber
\lefteqn{ {\cal D}(p,\Omega;m) \equiv 4\pi^2 d }\\
&& \times\left[ 
\frac{2\pi}{T} m_t^2 \delta_{0\,n} \delta^3(p) + 
\frac{1-\rm{e}^{
\left[-\right(p^2+\Omega^2+ m^2 \left)/(4m_t^2)\right]}}
        {p^2+\Omega^2+ m^2} \right]\; ,
\label{delta}
\end{eqnarray}
where $d= 12/(33-2N_f)$ and the ``Debye-mass'' is $m_{\rm D}^2 = \bar c T^2$,
$\bar c = 4\pi^2 d c$, $c= (N_c/3 + N_f/6)$, which is included in a manner
analogous to that in Ref.\cite{AAL89}.  There is no $T$-dependent mass in
$\Delta_G$.

The first term in Eq.~(\ref{delta}) is an integrable, infrared
singularity\cite{MN83} that generates long-range effects associated with
confinement~\cite{IRsing}.  The second term ensures that, neglecting
quantitatively unimportant logarithmic corrections, the propagator has the
correct perturbative behaviour at large spacelike arguments.  In this model
the large-distance, confining effects associated with $\delta_{0\,n}
\delta^3(p)$ are completely cancelled at small distances and one recovers the
perturbative result.  ${\cal D}(p,\Omega;m)$ has no Lehmann representation
and hence represents a confined particle, since this ensures the absence of
gluon production thresholds in ${\cal S}$-matrix elements describing
colour-singlet to singlet transitions.

Equation~(\ref{delta}) is an Ansatz based on $T=0$ studies augmented by
finite-$T$ perturbation theory.  Solving the gluon DSE at finite-$T$ would
provide additional insight but no such studies are currently available.
Without further information; for example, solving the coupled gluon-quark DSE
system or lattice estimates of the two-point gluon Schwinger function, we
cannot be sure that the $T$-dependence we have introduced provides a
qualitatively accurate representation of the temperature evolution of the
gluon propagator.  The paucity of relevant experimental data means that one
cannot presently use experiment to constrain this extrapolation.  This is the
primary source of uncertainty in our study and we therefore advocate a
cautious interpretation of our results near the transition temperature.

The model of Ref.~\cite{FR96} is recovered with two further simplifying
specifications.  The approximation
\begin{eqnarray}
\label{QEDWI}
Z_1=Z_2 & \;\;{\rm and}\;\; & Z_1^A=Z_2^A 
\end{eqnarray}
is used.  Gauge invariance in finite-$T$ QED entails these Ward identities.
In QCD the analogous identities are more complicated and this is only a
simplifying truncation, which has proven qualitatively and quantitatively
reliable in Landau gauge, $T=0$ studies.  In nonperturbative DSE
studies\cite{FR96,HW95} $Z_2$ and $Z_2^A$ are finite and $\approx 1$.

In addition, the rainbow approximation
\begin{equation}
\label{rainbow}
\Gamma_\mu(q,\omega_m;p,\omega_n) = \gamma_\mu
\end{equation}
is used.  In $T=0$ studies this has proven to be reliable in Landau gauge;
i.e., an efficacious phenomenology with a more sophisticated Ansatz only
requires a small quantitative modification of the parameters that
characterise the small-$k^2$ behaviour of the gluon
propagator~\cite{HRW94,HW91}.

Confinement can be investigated by studying the analytic properties of the
two-point Schwinger function of a given excitation.  The confinement test
proposed in Ref.~\cite{HRW94}, and used to good effect in Ref.~\cite{M95}, is
appropriate to this task. Consider 
$\Delta_{B_0}(x,0) \equiv (T/2\pi x)
\sum_{n=0}^\infty\,\Delta_{B_0}^n(x)$,
\begin{eqnarray}
\Delta_{B_0}^n(x) & \equiv & 
\frac{2}{\pi}\int_0^\infty\,dp\,p\,\sin(px)\,\sigma_{B_0}(p,\omega_n) \,.
\end{eqnarray}
For a free fermion of mass $m$: $\sigma_B(p,\omega_n)=
m/[\omega_n^2+p^2+m^2]$, hence $\Delta_B^n(x)= m
\,\exp(-x\,\sqrt{\omega_n^2+m^2})$.  This illustrates that the $n=0$ term
dominates the sum.

For a free fermion, $M(x;T)\equiv -
(d/dx)\,\ln\left|\Delta_{B_0}^0(x)\right|$ $=\surd(\pi^2 T^2 + m^2)$, a
finite constant.  Finite-$T$ effects become important for $T\sim m/\pi$.  In
DSE studies, $m$ is a mass-scale characteristic of dynamical mass generation:
$M_E^{u/d}\approx 300\,$MeV, hence we expect finite-$T$ effects to become
noticeable at $T\sim 100\,$MeV.

A Schwinger function with complex-conjugate poles, with $\Re(p^2)<0$
(timelike) and $\Im(p^2)\propto b$, does not have a Lehmann representation;
i.e., it represents a confined excitation with no associated asymptotic
state.  This is the nature of the quark Schwinger function obtained in
Ref.~\cite{FR96}.  In such cases $\Delta_{B_0}^0(x)$ has at least one zero.
Denote the position of the first zero in $\Delta_{B_0}^0(x)$ by $r_1$, and
define $\nu_0 \equiv 1/r_1$, then deconfinement is observed if, for some
$T=T_c$, $\nu_0(T_c)=0$; at this point thermal fluctuations have
overwhelmed the confinement scale-parameter and the poles have migrated to the
real-axis.

We solve the DSE for $S(p,\omega_n)$ numerically with
\begin{equation}
\begin{array}{ccc}
m_t = 0.69\;{\rm GeV}\,, & & m_R(\mu) = 1.1\;{\rm MeV}\;,
\end{array}
\end{equation}
which were fixed in Ref.~\cite{FR96} by requiring a best $\chi^2$-fit to a
range of $\pi$-meson observables at $T=0$.  We use $d=4/9$ in
Eq.~(\ref{delta}) and renormalise at $\mu=9.47\;$GeV~\cite{FR96}.  With these
choices our current study has no free parameters.  We use $\bar\Lambda/\mu =
1$, $|\omega_{n_{\rm max}}|/|\vec{p}|_{\rm max} \approx 1$, $n_{|p|}=64$
points in the $|p|$-array. Our results are cutoff independent.  At $T=5\,$MeV
the results in Table~I of Ref.~\cite{FR96} are reproduced to within 6\%.

At finite-$T$ the pion mass is given by~\cite{FR96}
\begin{equation}
\label{pimass}
m_\pi^2\,N_\pi^2 = \langle m_R(\mu) \, (\bar q q)_\mu \rangle_\pi
\end{equation}
with ($\int d\mu_T \equiv (2 T)/(\pi^2)\,\sum_{n=0}^\infty\,
        \int_0^{\bar \Lambda}\,dp\,p^2$)
\begin{eqnarray}
\nonumber
\lefteqn{\langle m_R(\mu) \, (\bar q q)_\mu \rangle_\pi  \equiv}\\
& & 4 N_c\int d\mu_T\,B_0\,\left(
        \sigma_{B_0} - 
        B_0\,\left[
        \omega_n^2 \sigma_C^2 +
        p^2 \sigma_A^2 + 
        \sigma_B^2 \right] \right)\;,
\label{mqbq}
\end{eqnarray}
which vanishes linearly with $m_R(\mu)$, and
\begin{eqnarray}
\nonumber
\lefteqn{N_\pi^2  =  N_c \int d\mu_T\,
B_0^2\,
\left\{\sigma_A^2 - 2 \left[ 
        \omega_n^2\sigma_C\sigma_C^\prime + 
        p^2 \sigma_A\sigma_A^\prime \right. \right.} \\
& & \nonumber
 + \left. \sigma_B\sigma_B^\prime\right]
   - \case{4}{3}\,p^2\,\left(
        \left[ \omega_n^2\left(\sigma_C\sigma_C^{\prime\prime} -
        (\sigma_C^\prime)^2\right) \right. \right.\\
& & + \left. \left. \left.
        p^2\left(\sigma_A\sigma_A^{\prime\prime} -
        (\sigma_A^\prime)^2\right) + 
        \sigma_B\sigma_B^{\prime\prime} -
        (\sigma_B^\prime)^2 \right] \right) \right\}\;,
\label{npisq}
\end{eqnarray}
with $\sigma^\prime_B \equiv \partial\sigma_B(p^2,\omega_n)/\partial p^2$
etc., is the canonical normalisation constant for the Bethe-Salpeter
amplitude in ladder approximation.  The pion decay constant is obtained from
\begin{eqnarray}
\nonumber
\lefteqn{f_\pi\,N_\pi   =}\\
&&   2 N_c\int d\mu_T\,B_0\,
\left\{ \sigma_A \sigma_B + 
        \case{2}{3} p^2 \left(\sigma_A^\prime \sigma_B - 
                                \sigma_A \sigma_B^\prime\right)\right\}\;.
\label{fpi}
\end{eqnarray}
Equation~(\ref{pimass}) provides an accurate estimate of the mass obtained in
solving the dressed-ladder Bethe-Salpeter equation for the pion~\cite{FR96}.
Equations (\ref{mqbq})-(\ref{fpi}) neglect corrections of order $m_\pi^2$,
which are unimportant ($<3$\%) for all values of $T$; i.e., $m_\pi$ is
dominated by a linear response to $m_R$ at all $T$.

In deriving these formulae one assumes $\Gamma_\pi = i\gamma_5\,B_0$, where
$\Gamma_\pi$ is the pion Bethe-Salpeter amplitude, which is a manifestation of
Goldstone's theorem in the DSEs\cite{DS79BRS96}.  Since
$B_0$ is nonzero only if chiral symmetry is dynamically broken, these
formulae are not valid above any chiral symmetry restoration temperature,
$T_c^\chi$.

We employ the simplest order parameter for DCSB:
\begin{equation}
\chi \equiv B_0(p=0,\omega_0)\,.
\end{equation}

\begin{figure}[h,t]
  \centering{\
     \epsfig{figure=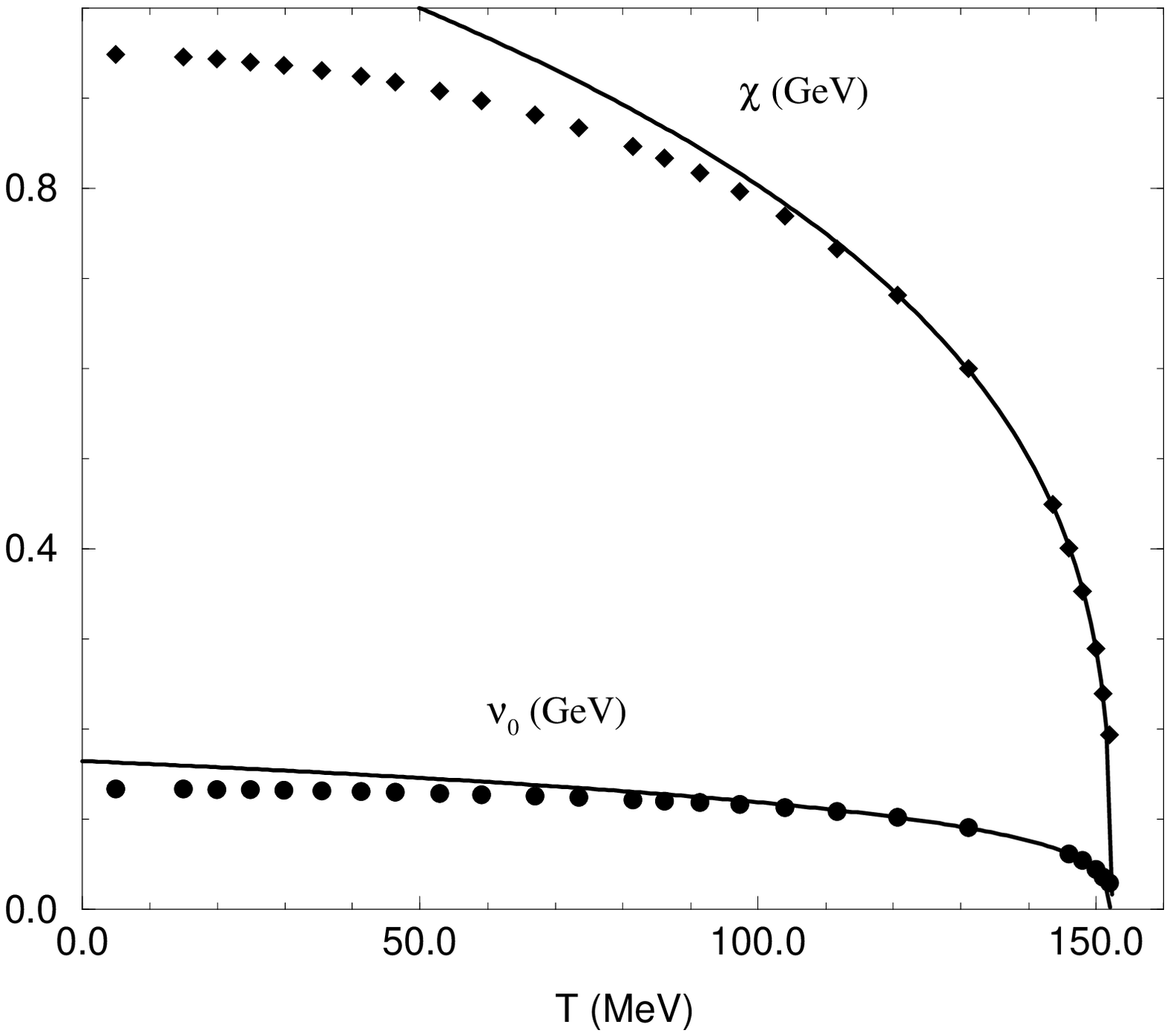,height=13.5cm,rheight=13.5cm}  }
\caption{The parameters for the curves, $\chi(T)$ (diamonds) and $\nu_0(T)$
(circles), are presented in Eq.~(\protect\ref{critparam}).
\label{fitcrit}}
\end{figure}

We plot $\chi(T)$ and $\nu_0(T)$ in Fig.~\ref{fitcrit}.  The curves, fitted
on $120\leq T\,({\rm MeV}) \leq 150$, are of the form $\alpha\,
(1-T/T_c)^\beta$ with
\begin{equation}
\label{critparam}
\begin{array}{lll}
\alpha_\chi = 1.1\,{\rm GeV}\,, 
        & \beta_\chi = 0.33\,, & T_c^\chi = 150\,{\rm MeV}\\
\alpha_{\nu_0} = 0.16\,{\rm GeV}\,, & \beta_{\nu_0}= 0.30\,, 
                & T_c^{\nu_0} = 150\,{\rm MeV}\,.
\end{array}
\end{equation}
The transitions are coincident and have the same critical exponent, within
errors: $\sim 10$\%.

The massive quark is not deconfined at the same temperature. However, the
behaviour of $\nu_m$ undergoes a qualitative change at
$T_c^{\nu_0}=T_c^\chi$, where no contribution to $B(p,\omega_n)$ remains
that does not vanish as $m_R \to 0$.  For $T\leq T_c^{\chi}$:
$\alpha_{\nu_m} = 0.15\,{\rm GeV}$, $\beta_{\nu_m}= 0.23$,
$T_c^{\nu_m} = 160\,{\rm MeV}$; for $T>T_c^{\chi}$:
$\alpha_{\nu_m} = 0.26\,{\rm GeV}$, $\beta_{\nu_m}= 0.71$,
$T_c^{\nu_m} = 180\,{\rm MeV}$.  For $T>T_c^{\chi}$, a fit with
$\beta_{\nu_m}= 0.23$ has a five-times larger standard deviation.  The
$u$/$d$-quark in this model is therefore deconfined at $T\sim 180$~MeV,
suggesting that the deconfinement temperature increases with current-quark
mass.

\begin{figure}[h,t]
  \centering{\
    \epsfig{figure=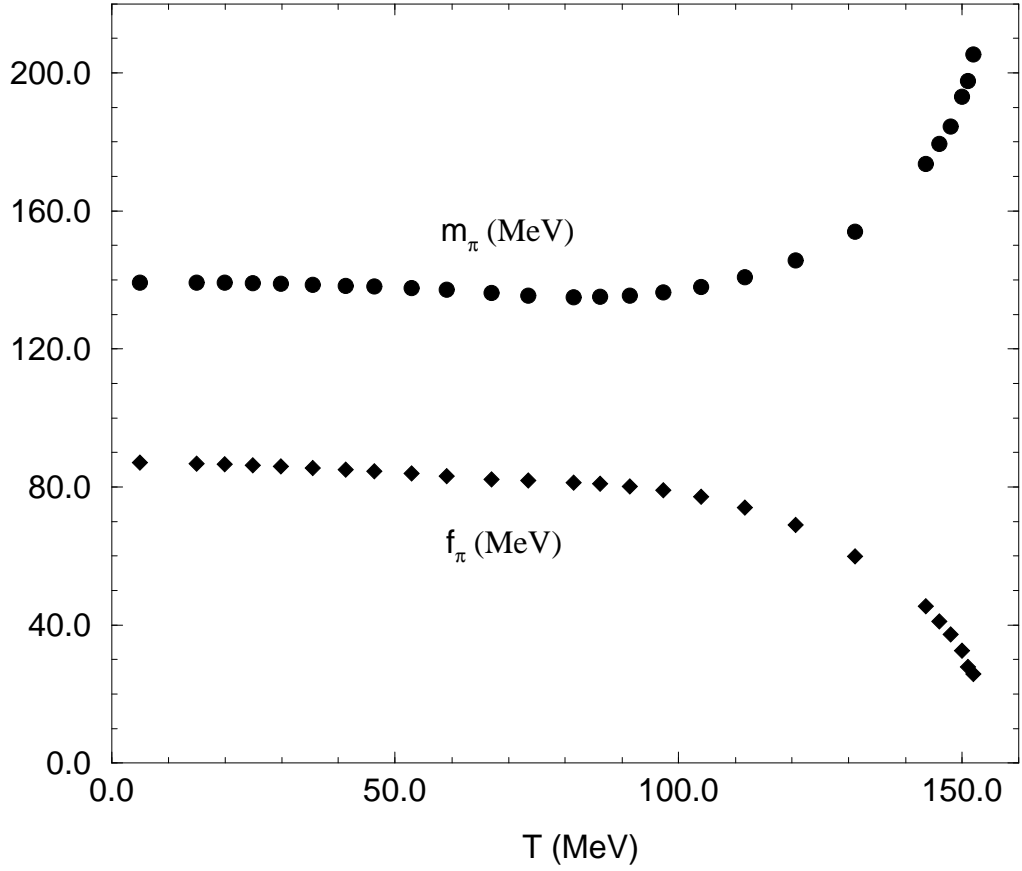,height=13.5cm,rheight=13.5cm}  }
\caption{$f_\pi(T)$ (diamonds) and $m_\pi(T)$ (circles) plotted against $T$. 
See the discussion around Eq.~(\protect\ref{critfpimpi}).
\label{figfpimpi}}
\end{figure}

We find that, on $120\leq T ({\rm MeV})\leq 150$, $N_\pi^2$, $N_\pi\,f_\pi$
and $\langle m_R(\mu) (\bar q q)_\mu\rangle$ are described by $\alpha\,
(1-T/T_c)^\beta$, with $T_c = 150\,$MeV in each case and
\begin{equation}
\label{critnpifpimqbq}
\begin{array}{rclrcl}
\alpha_{N_\pi^2} & = & (0.18\,{\rm GeV})^2\,, & \beta_{N_\pi^2} & = & 1.1\,, \\
\alpha_{f_\pi N_\pi} & = & (0.15\,{\rm GeV})^2\,, & \beta_{f_\pi N_\pi} & = &
0.93\,, \\
\alpha_{\langle m_R (\bar q q)\rangle} & = & (0.15\,{\rm GeV})^4\,, &
\beta_{\langle m_R (\bar q q)\rangle}& = & 0.92\,.
\end{array}
\end{equation}
In order to avoid round-off errors associated with division near $T=T_c$ we
used these results to determine 
\begin{equation}
\label{critfpimpi}
\begin{array}{rclrcl}
\alpha_{m_\pi} & = & 0.12\,{\rm GeV}\,, & \beta_{m_\pi} & = & -0.11\,,\\
\alpha_{f_\pi} & = & 0.12\,{\rm GeV}\,, & \beta_{f_\pi}& = & 0.36\,,
\end{array}
\end{equation}
with $T_c^{m_\pi} = 150\,{\rm MeV}\,=T_c^{f_\pi}$.  We plot $m_\pi(T)$ and
$f_\pi(T)$ in Fig.~\ref{figfpimpi}.

We explored the finite-$T$ properties of a renormalisable, confining,
DSE-model of QCD~\cite{FR96}.  Introducing an order parameter for confinement
we found that, in the chiral limit, a deconfinement transition at $T\approx
150\,$MeV is accompanied by the coincident restoration of chiral symmetry.
The single model parameter, fixed at $T=0$, is appropriate for two-flavour
QCD: the transitions are second order and are not described by mean-field
critical exponents.  Similar results have been obtained in recent numerical
simulations of lattice-QCD at finite-$T$\cite{FK95}.

$f_\pi$ and $m_\pi$ are weakly sensitive to $T$ for $T< 0.7\,T_c^\chi$.
However, as $T$ approaches $T_c^\chi$, the mass eigen value in the pion
Bethe-Salpeter equation moves to increasingly larger values, as thermal
fluctuations overwhelm attraction in the channel, until at $T=T_c^\chi$ there
is no solution and $f_\pi \to 0$.  This means that the pion-pole contribution
to the four-point, quark-antiquark correlation function disappears; i.e.,
there is no quark-antiquark pseudoscalar bound state for $T>T_c^\chi$.  This
may have important consequences for a wide range of physical
observables~\cite{Betal95}, if borne out by improved studies; e.g., this
$T$-dependence of $f_\pi$ and $m_\pi$ would lead to a 20\% reduction in the
$\pi \to \mu \nu_\mu$ decay widths at $T\approx 0.9\,T_c^\chi$.


DB is grateful for the hospitality of the Physics Division at ANL during a
visit in which this work was conceived.  CDR is grateful for the hospitality
of the members of the MPG at the University of Rostock during two visits in
which part of this research was conducted and for a stipend from the Max
Planck Gesellschaft, which supported these visits.  This work was supported
by the US Department of Energy, Nuclear Physics Division, under contract
number W-31-109-ENG-38.  The calculations described herein were carried out
using a grant of computer time and the resources of the National Energy
Research Supercomputer Center.


\end{document}